\documentstyle[aps]{revtex}

\newcommand{\infig}[2]{\begin{center}\mbox{\epsfxsize #2
\epsfbox{#1}}\end{center}}

\newtheorem{theorem}{Theorem}
\newtheorem{acknowledgement}[theorem]{Acknowledgement}

\begin{document}
\input {epsf}
\title{Long distance entanglement based quantum key distribution}
\author{Gr\'{e}goire Ribordy, J\"{u}rgen Brendel$^{\dagger }$, Jean-Daniel Gautier,
Nicolas Gisin, and Hugo Zbinden}
\address{Gap-Optique, Universit\'{e} de Gen\`{e}ve, 20 rue de\\
l'Ecole-de-M\'{e}decine, 1211 Gen\`{e}ve 4, Switzerland\\
Email: gregoire.ribordy@physics.unige.ch\\
$^{\dagger }$Present address: Luciol Instruments SA, 31 Chemin de la\\
Vuarpilli\`{e}re, 1260 Nyon, Switzerland}
\date{\today}
\maketitle

\begin{abstract}
A detailled analysis of quantum key distribution employing entangled states
is presented. We tested a system based on photon pairs entangled in
energy-time optimized for long distance transmission. It is based on a
Franson type set-up for monitoring quantum correlations, and uses a protocol
analogous to BB84. Passive state preparation is implemented by polarization
multiplexing in the interferometers. We distributed a sifted key of 0.4
Mbits at a raw rate of 134 Hz and with an error rate of 8.6\% over a
distance of 8.5 kilometers. We discuss thoroughly the noise sources and
practical difficulties associated with entangled states systems. Finally the
level of security offered by this system is assessed and compared with that
of faint laser pulses systems.
\end{abstract}

\section{Introduction}

Quantum key distribution (QKD), the most advanced application of the new
field of quantum information theory, offers the possibility for two remote
parties - Alice and Bob - to exchange a secret key without meeting or
resorting to the services of a courier. This key can in turn be used to
implement a secure encryption algorithm, such as the ``one-time pad'', in
order to establish a confidential communication link. In principle, the
security of QKD relies on the laws of quantum physics, although this claim
must be somewhat softened because of the lack of ideal components - in
particular the photon source and the detectors.

After the first proposal by Bennett and Brassard \cite{bb84}, various
systems of QKD have been introduced and tested by groups around the world
(see \cite{townsend98,hughes00,mérolla99,ribordy00} for recent experiments).
Until recently, all QKD experiments relied on strongly attenuated laser
pulses, as an approximation to single photons, because of the lack of
appropriate sources for such states. Although this solution is the simplest
from an experimental point of view, it suffers from two important drawbacks.
First, the fact that a fraction of the pulses contains more than one photon
constitutes a vulnerability to certain eavesdropping strategies. Second, the
maximum transmission distance is reduced, because of the fact that most of
the pulses are actually empty. Both points are discussed in more detail
below.

Ekert proposed in 1992 a protocol utilizing entangled states for QKD \cite
{ekert91}. Photon pair sources making use of parametric downconversion are
relatively simple and flexible. They have been used for several years and
were exploited for example for tests of Bell inequalities \cite
{tapster94,tittel98,weihs98}. These experiments demonstrated that
entanglement of photon pairs can be preserved over long distances in optical
fibers, and could thus allow the implementation of QKD.

Recently, research groups performed the first entangled photons pairs QKD
experiments \cite{tittel00,naik00,jennewein00}. Both Naik \cite{naik00} and
Jennewein \cite{jennewein00} chose to use photons at a wavelength of 702 nm,
entangled in polarization for their investigations. Although this choice is
appropriate for free space QKD, it prevents any transmission over a distance
of more than a few kilometers in optical fibers. Polarization entanglement
is indeed not very robust to decoherence, and attenuation at this wavelength
is rather high in optical fibers. Tittel {\it et al.} used photons pairs
correlated in energy and time and with a wavelength where the attenuation in
fibers is low, but their actual implementation was not optimized for long
distance transmission \cite{tittel00}.

In this paper, we present the first system for QKD with entangled photon
pairs exploiting an asymmetric configuration optimized for long distance
operation. In addition, we believe that it offers a particularly high level
of security. We introduce first the principle of our system, then discuss
experimental results obtained under laboratory conditions. Finally, we
compare it with other experiments and evaluate its advantages and drawbacks,
before concluding.

\section{Principle of the QKD system}

When designing a QKD system where photons are exchanged between Alice and
Bob, one must first choose on which property to encode the qubits values.
Although polarization is a straightforward choice, it is not the most
appropriate one when transmitting photon pairs over optical fibers. The
intrinsic birefringence of these fibers, also known as polarization mode
dispersion, associated with the large spectral width (typically 5 nm FWHM at
800 nm) of the down-converted photons yields rapid depolarization.
Considering that such photons have typically a coherence time of the order
of 1 ps, and that standard telecommunications fibers exhibit a polarization
mode dispersion of $0.2$ ps\thinspace /$\,\surd $km, one sees that the
polarization mode separation already becomes substantial after a few
kilometers. This fact indicates that polarization is not robust enough for
long distance QKD in fibers when using photon pairs. A solution is therefore
to encode the values of the qubits on the phase of the photons. In addition,
previous experiments demonstrated that the polarization transformation
induced by an installed optical fiber changes sometimes abruptly. An active
polarization alignment system is consequently necessary to compensate these
fluctuations.

A second important parameter for a QKD system is the wavelength of the
photons. Two opposite factors influence this choice. On the one hand, the
attenuation in optical fibers decreases with an increase of the wavelength
from 2 dB\thinspace /\thinspace km at 800 nm to a local minimum of 0.35
dB\thinspace /\thinspace km at 1300 nm and an absolute minimum of 0.25
dB\thinspace /\thinspace km at 1550 nm. On the other hand, photons with
lower energy - or longer wavelength - tend to be more difficult to detect.
Below 900 nm, one typically uses commercial modules built around a silicon
avalanche photodiode (Si APD) biased above breakdown. They offer good
quantum detection efficiency (typically 50\%), low noise count rate (100
Hz), and easy operation. In the so-called second telecom window, germanium
avalanche photodiodes (Ge APD) can be used. Their performance is not as good
as that of Si APD's and they require liquid nitrogen cooling. Finally, only
indium gallium arsenide avalanche photodiodes (InGaAs APD) exhibit
sufficient detection efficiency in the third telecom window around 1550 nm.
They have the same drawbacks as Ge APD's, but also require gated operation
to yield low enough dark counting rates. Taking into account these factors,
one can conclude that up to a few kilometers, 800 nm is a good choice. In
addition, beyond 30 to 40 km, the only real possibility is to operate the
system at 1550 nm, because fiber attenuation becomes really critical.

\subsection{QKD Protocol}

\begin{figure}
\infig{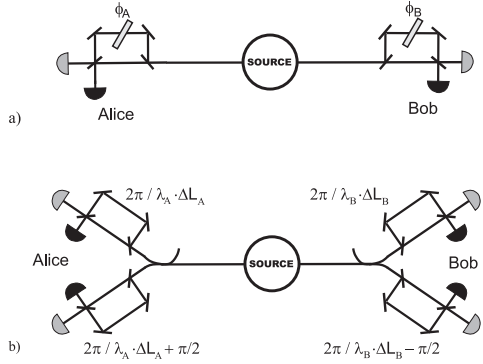}{0.95\columnwidth}
\caption{a) Franson type arrangement for generating non-local quantum
correlations with photon pairs entangled in energy time. b) Implemenation of
the double measurement basis with four interferometers.}
\label{FigInterf}
\end{figure}

Our system is based on a Franson arrangement \cite{franson89}. It exploits
photon pairs entangled in energy-time, where both the sum of the energy and
of the momenta of the down-converted photons equal those of the pump photon.
A source located between Alice and Bob generates such pairs, which are split
at its output (see Fig. \ref{FigInterf} a). One photon is sent to each party
down quantum channels. Both Alice and Bob possess an unbalanced Mach-Zehnder
interferometer, with photon counting detectors connected at its outputs.
When considering a given photon pair, four events can yield coincidences
between one detector at Alice's and one at Bob's. First, the photons can
both propagate through the short arms of the interferometers. Then, one can
take the long arm at Alice's, while the other takes the short one at Bob's.
The opposite is also possible. Finally, both photons can propagate through
the long arms. When the path differences of the interferometers are matched
within a fraction of the coherence length of the down-converted photons, the
short - short and the long - long processes are indistinguishable and yield
thus two-photons interference, provided that the coherence length of the
pump photons is longer than the path difference. If one monitors these
coincidences as a function of time, three peaks appear. The central one is
constituted by the interfering short-short and long-long events. It can be
separated from non-interfering ones, by placing a window discriminator. Only
interfering processes will be considered below.

We implemented a protocol analogous to BB84. Ekert {\it et al.} showed in
\cite{ekert92} that the probabilities for Alice and Bob to get correlated
counts (the photons choose the same port at Alice's and Bob's) and
anticorrelated counts (they choose different ports) are given by

\begin{equation}
P_{correlation}=P\left( A=0;B=0\right) +P\left( A=1;B=1\right) =\frac{1}{2}%
\left[ 1+\cos \left( \phi _{A}+\phi _{B}\right) \right]  \label{pcorr}
\end{equation}

\begin{equation}
P_{anticorrelation}=P\left( A=0;B=1\right) +P\left( A=1;B=0\right) =\frac{1}{%
2}\left[ 1-\cos \left( \phi _{A}+\phi _{B}\right) \right]  \label{panticorr}
\end{equation}

where Alice's phase $\phi _{A}$ and Bob's phase $\phi _{B}$ can be set
independently in each interferometer. The results of Alice's and Bob's
measurements are represented by $A$ and $B$. They can take values of 0 or 1
depending on the detector that registered the count. One sees that if the
sum of the phases is equal to 0, $P_{correlation}=1$ and $%
P_{anticorrelation}=0$. In this case, Alice can deduce that whenever she
gets a count in one detector, Bob will also get one in the associated
detector. If both Alice and Bob set their phases to 0, they can exchange a
key by associating a bit value to each detector. However, if they want their
system to be secure against eavesdropping attempts, they must implement a
second measurement basis. This can for example be done by adding a second
interferometer to their systems (see Fig. \ref{FigInterf}\ b). Now, when
reaching an analyzer, a photon chooses randomly to go to one or the other
interferometer. The phase difference between Alice's interferometers is set
to $\pi /2$, whereas the one between Bob's to $-\pi /2$. If both photons of
a pair go to associated interferometers, the sum of the phase they
experience is 0. We obtain again the correlated outcomes discussed above. On
the contrary, if they go to different interferometers, the sum is $\pm \pi
/2 $. In this case, one finds that $P_{correlation}=%
{\frac12}%
$ and $P_{anticorrelation}=%
{\frac12}%
$. Alice's and Bob's outcomes are then not correlated at all. They perform
incompatible measurements. After exchanging a sequence of pairs, the parties
must of course go through the conventional steps of key distillation, as in
any QKD system: key sifting, error correction and privacy amplification \cite
{bennett92}.

\subsection{Photon Pairs Configuration}

\begin{figure}
\infig{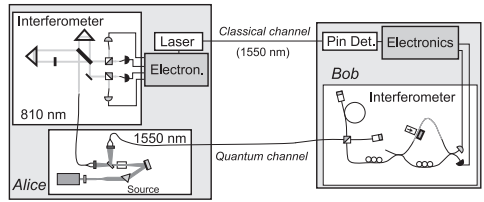}{0.95\columnwidth}
\caption{Asymmetric system for quantum key distribution utilizing photon
pairs.}
\label{FigSystème}
\end{figure}

Let us now discuss the choice of wavelength for the photons of the pairs. As
mentioned earlier, when transmitting photons over long distances, one should
select a wavelength of 1550 nm to minimize fiber attenuation. However,
detectors sensitive to such photons require gated operations in order to
keep the dark counting rates low. Therefore, we selected an asymmetrical
configuration where only the photon travelling to Bob has this wavelength,
while the one travelling to Alice has a wavelength below 900 nm. She can
consequently use free running Si APD detectors. Whenever she gets a click,
she sends a classical signal to Bob to warn him to gate his detectors. The
source is located very close to Alice's interferometers, to keep fiber
attenuation negligible (see Fig. \ref{FigSystème}). One should note that in
such an asymmetrical configuration, the losses in Alice's apparatus seem to
be unimportant. When a photon gets lost in Alice's analyzer, she does not
send a classical signal to Bob, who in turns does not gate his detectors.
Such an event can thus not yield a false count through detector noise. A
second possibility is to utilize one photon of the pair simply to generate a
trigger signal, indicating the presence of the other one. This solution is
not optimal. The second photon must indeed be sent through a preparation
device featuring attenuation, which will reduce its probability to be
detected by Bob.

Systems using pairs of photons entangled in energy-time are more sensitive
to chromatic dispersion spreading in the transmission line than faint pulse
set-ups, because of the relatively large spectral width of the pairs.
Indeed, interfering events are discriminated from non-interfering ones by
timing information. Spreading of the photons between Alice and Bob induced
by chromatic dispersion must thus be kept to a minimum. For example,
assuming a spectral width of 6 nm, an impulsion launched in a standard
singlemode fiber featuring a typical dispersion coefficient of $18$ ps$\,$nm$%
^{-1}\,$km$^{-1}$ at 1550 nm would spread to 1 ns after 10 km of fiber. This
effect can be avoided by using dispersion shifted fibers (DS-fibers) with
their dispersion minimum close to the down-converted wavelength. It is also
possible to compensate dispersion (see \cite{DS-comp}), although it implies
additional attenuation.

\subsection{Characterizing the System}

In order to characterize our new system and assess its advantages over other
set-ups, we introduce in this section the equations expressing the quantum
bit error rate (QBER) and the sifted key distribution rate.

In principle, when an eavesdropper - Eve - performs a measurement on a qubit
exchanged between Alice and Bob, she induces a perturbation with non-zero
probability, yielding errors in the bit sequence. These discrepancies reveal
her presence. Nevertheless in practical systems, errors also happen because
of experimental imperfections. One can quantify the frequency of these
errors as the probability of getting a false count over the total
probability of getting a count (see Eq. (\ref{QBER1})). In the limit of low
error probability, this ratio can be approximated by the probability of
getting an incorrect count over the probability of getting a correct one. As
discussed above, Bob's detectors are operated in gated mode and these
probabilities must thus be calculated per gate. In addition, we will
consider only the cases where compatible bases are selected by Alice and Bob.

\begin{equation}
QBER=\frac{\text{prob. incorrect count}}{\text{prob. incorrect + correct
count}}\approx \frac{\text{prob. incorrect count}}{\text{prob. correct count}%
}  \label{QBER1}
\end{equation}

The correct count probability is expressed as the product of several terms.
The first one is $\mu $, the probability of having a photon leaving the
source in the direction of Bob whenever Alice detects a photon and sends a
classical pulse. Then come the probabilities $T_{L}$ and $T_{B}$ for this
photon to be transmitted respectively by the fiber link and by Bob's
apparatus. The next factor, $q_{interf}$, is equal to $%
{\frac12}%
$ in our system and takes into account the fact that only half of the
photons will actually yield interfering events that can be used to generate
the key. The factor $\eta _{D}$ represents the quantum detection efficiency
of Bob's detectors. Finally, the term $q_{basis}$ accounts for the cases
where Alice and Bob perform incompatible measurements. It is equal to $%
{\frac12}%
$ for symmetrical basis choice.

\begin{equation}
P_{correct}=\mu \cdot T_{L}\cdot T_{B}\cdot q_{interf}\cdot \eta _{D}\cdot
q_{basis}  \label{pcorrect}
\end{equation}

The probability of getting a false count per gate can be thought of as the
sum of three terms. It can arise first through a detector error. Each one of
our four detectors can register a noise count. It will yield an error in
25\% of the cases, a correct bit also in 25\% of the cases and an
incompatible measurement in the remaining 50\%. It is thus accounted for by
the probability $p_{cs}$ ($=4\cdot p_{cs}\cdot
{\frac14}%
$).

\begin{equation}
P_{incorrect}=p_{cs}+\mu \cdot T_{L}\cdot T_{B}\cdot q_{interf}\cdot \eta
_{D}\cdot q_{basis}\cdot p_{opt}+\nu \cdot T_{L}\cdot T_{B}\cdot
q_{interf}\cdot \eta _{D}\cdot q_{basis}\cdot q_{acc}  \label{pincorrect}
\end{equation}

The second term corresponds to the cases where, because of imperfect phase
alignment of the interferometers, a photon chooses the wrong output port of
the interferometer. It is given by the product of the probability of getting
a correct count multiplied by the probability $p_{opt}$ for the photon to
choose the wrong port. In an interferometric system, it stems from non-unity
visibility $V$, and is given by the following equation.

\begin{equation}
p_{opt}=\frac{1-V}{2}  \label{popt}
\end{equation}

Finally, the last term takes into account the probability of getting a count
from an accidental coincidence. It is given by the product of the
probability $\nu $ of having an uncorrelated photon within a gate, with the
probability for this photon to reach Bob's system, and be detected in the
compatible basis. Because of the fact that it is not correlated with
Alice's, this photon will choose the output randomly and yield a false count
in 50\% of the cases, and a correct one also in 50\%. This is accounted for
by the factor $q_{acc}$, which is equal to $%
{\frac12}%
$.

These three components can be separated into three QBER contributions as in
Eq. (\ref{QBER2}). These formula are general and thus still valid for other
systems.
\begin{equation}
QBER=QBER_{det}+QBER_{opt}+QBER_{acc}  \label{QBER2}
\end{equation}

\begin{equation}
QBER_{det}=\frac{p_{cs}}{\mu \cdot T_{L}\cdot T_{B}\cdot q_{interf}\cdot
\eta _{D}\cdot q_{basis}}  \label{QBERdet}
\end{equation}

\begin{equation}
QBER_{opt}=\frac{\mu \cdot T_{L}\cdot T_{B}\cdot q_{interf}\cdot \eta
_{D}\cdot q_{basis}\cdot p_{opt}}{\mu \cdot T_{L}\cdot T_{B}\cdot
q_{interf}\cdot \eta _{D}\cdot q_{basis}}=p_{opt}  \label{QBERopt}
\end{equation}

\begin{equation}
QBER_{acc}=\frac{\nu \cdot T_{L}\cdot T_{B}\cdot q_{interf}\cdot \eta
_{D}\cdot q_{basis}\cdot q_{acc}}{\mu \cdot T_{L}\cdot T_{B}\cdot
q_{interf}\cdot \eta _{D}\cdot q_{basis}}=q_{acc}\frac{\nu }{\mu }
\label{QBERacc}
\end{equation}

One should note that if the basis choice was implemented actively, only two
out of the four detectors at Bob's would be gated for a given bit. This
implies that both $QBER_{det}$ and $QBER_{acc}$ would be reduced by a factor
of two. In principle, active switching ensures thus a gain of a factor 2 in $%
QBER_{det}$, corresponding to approximately 10 km of transmission distance
at 1550 nm. In practice, this is not true because of the additional losses
induced by devices used to perform active bases choices (Pockel cells or
LiNbO$_{3}$-phase modulators for example).

When the length of the fiber link is increased, $T_{L}$ decreases. The
probability of getting a right count is reduced, while the probability of
registering a dark count remains constant and $QBER_{det}$ thus increases.
On the other hand, as they do not depend on $T_{L}$, both $QBER_{opt}$ and $%
QBER_{acc}$ remain unchanged. When exchanging key material over long
distances, $QBER_{det}$ becomes consequently the main contribution and sets
an ultimate limit on the span. In order to maximize the distance, one should
clearly choose the best detectors available, and maximize the correct count
probability. In systems exploiting faint laser pulses, it is essential that
the multiphoton pulse probability is low to ensure security. In this case,
one selects for $\mu $ a value well below unity, which reduces the correct
count probability. A given QBER is thus reached for a shorter transmission
distance. Setting this parameter to 0.1 - a typical value - instead of 1 has
the same effect on $QBER_{det}$ as adding fiber attenuation of 10 dB,
corresponding to a distance of about 40 km at 1550 nm. One sees clearly a
first advantage of using photon pairs instead of faint laser pulses. This
issue is discussed in more details in Sec. \ref{discus}.

Unfortunately, additional factors reduce this advantage. Comparing the
predicted performance of our photons pairs system with that of a well tested
faint pulses system like our ``plug \& play'' set-up \cite{ribordy00}, we
see that the ratio of $QBER_{det}$ for a given transmission distance is in
theory equal to

\begin{equation}
\frac{QBER_{det}^{p\&p}}{QBER_{det}}=\frac{\mu \cdot q_{interf}}{2\cdot \mu
^{p\&p}\cdot q_{interf}^{p\&p}}=\frac{5}{2}  \label{QBERppQBER}
\end{equation}

This results is obtained by setting $q_{interf}^{p\&p}=1$ and $\mu
^{p\&p}=0.1$ for the ``plug \& play'' and $q_{interf}=%
{\frac12}%
$ and $\mu =1$ for our photon pairs system. The factor 2 in the denominator
comes from the fact that active basis selection is performed with the
''plug\&play'' system. The other factors are assumed to be identical and
they just cancel out. This means that our new system should be able to
handle 4 dB of additional fiber attenuation, corresponding to approximately
16 km at 1550 nm. However, one should note that photon pairs systems suffer
from an additional contribution to their error rate - $QBER_{acc}$ - which
somehow reduces this advantage. Although it is important, this span increase
would not revolutionize the potential applications of QKD over optical
fibers.

Finally, it is possible to estimate the actual raw key creation rate (after
sifting, but before distillation) by multiplying the probability of getting
a right count by the counting rate registered by Alice.

\begin{equation}
R_{raw}=f_{Alice}\cdot P_{correct}  \label{Rraw}
\end{equation}

The quantity $f_{Alice}$ represents the repetition frequency and $%
P_{correct} $ is given by Eq. (\ref{pcorrect}). One can then apply
correction factors to estimate the distilled key rate \cite{lütkenhaus99}.

\section{Implementation of the system}

Now that the principles of QKD using photon pairs entangled in energy-time
have been discussed, we can consider the actual implementation of the
system. It consists of four basic subsystems: the photon pair source,
Alice's interferometer, Bob's interferometer, and the classical channel
(Fig. \ref{FigSystème}). We also discuss the procedure used to measure and
adjust the path differences of the interferometers.

\subsection{The Photon Pair Source}

\begin{figure}
\infig{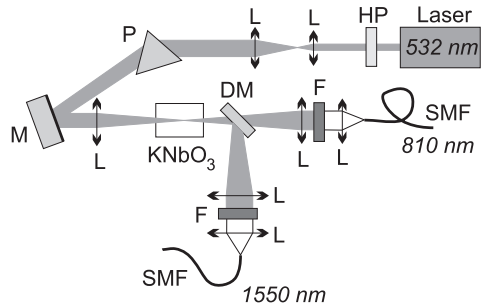}{0.95\columnwidth}
\caption{Schematic diagram of the photon pair source (HW: half-wave plate,
L: lens, P: dispersive prism, M: metallic mirror, DM: dichroic mirror, F:
filter, SMF: single-mode fiber)}
\label{FigSource}
\end{figure}

The source is basically made up of a pump laser, a beam shaping and delivery
optical system, a non-linear crystal and two optical collection systems (see
Fig. \ref{FigSource}). It is built with bulk optics. The pump laser is a
GCL-100-S frequency doubled-YAG laser manufactured by Crystalaser. It emits
100 mW of single mode light at 532 nm. Its spectral width is narrower than
10 kHz. This corresponds to a coherence length of about 30 km for the pump
photons, and yields in turn a high visibility for the two-photons
interference. Its frequency stability was verified to be better than 50 MHz
per 10 minutes. This is an important parameter since the wavelength of the
pump photons controls the wavelengths of the down-converted photons. These
must remain stable during a key distribution session, because they determine
the relative phases the photons experience in the interferometers.
The collimated beam passes first through a half-wave plate, which rotates
its linear polarization state to horizontal. It goes then through a
keplerian beam expander ($\times $2). It passes then a dispersive prism and
a Schott BG39 band pass filter (T = 98\% at 532 nm, and T = 10$^{-4}$ at
1064 nm), in order to remove any infrared light that might mask actual
photon pairs. Both of these components are aligned so that the angle between
their surfaces and the incident beam is close to the Brewster angle, in
order to minimize pump power loss by partial reflection. The beam is then
reflected by a metallic mirror before going through a pinhole, which
complements our simple monochromator. It is then focused on the KNbO3
non-linear crystal through a biconvex achromat with 100 mm focal length. The
crystal measures 3 ($\phi $-plane) x 4 ($\theta $-plane) x 10 mm$^{3}$. It
is cut with a $\theta $-angle of 22.95%
${{}^\circ}$%
and allows collinear down-conversion at 810 nm and 1550 nm when kept at room
temperature and illuminated normally with a pump at 532 nm. Its first face
is covered with AR coating for 532 nm, while the second one has AR coating
for 810 and 1550 nm. The crystal can be slightly rotated ($\pm \ 5%
{{}^\circ}%
$) to tune the pump incidence angle. This parameter is used to adjust the
down-converted wavelengths. The down-converted beams are then split by a
dichroic mirror aligned at $45%
{{}^\circ}%
$ incidence. The photons at 810 nm experience a transmission coefficient of
approximately 80\%, while the 1550 nm ones experience a reflection
coefficient of more than 98\%. The short wavelength beam is then collimated
by a biconvex achromat with a focal length of 150 mm. A set of two uncoated
filters is then used to block off the pump light. One should avoid
fluorescence in this process, in order to minimize the probability of
recording coincidences from uncorrelated photons. This is achieved by using
first a low fluorescence Schott KV550 long pass filter (T=20\% at 532 nm) to
reduce the pump intensity, before blocking it with a Schott RG715 long pass
filter. The 810 nm photons are then focused onto the core of a singlemode
fiber (cutoff wavelength
\mbox{$<$}%
780 nm, mode field diameter = 5.5 $\mu $m) by a collimator (focal length =
11 mm) with FC/PC receptacle.

After being reflected by the dichroic mirror, the 1550 nm-beam is collimated
by a biconvex achromat with focal length of 75 mm. The pump beam is then
removed by a coated silicon long wave pass filter (5\% cut-on at 1050 nm),
offering a transmission coefficient at 1550 nm close to 100\%. The
down-converted beam is then focused onto the core of a singlemode fiber
(cutoff wavelength
\mbox{$<$}%
1260 nm, mode field diameter = 10.5 $\mu $m), through an identical fiber
collimator as the 810 nm beam.

As discussed above, the probability $\mu $ of having one photon at 1550 nm
leaving the source, knowing that there was one at 810 nm must be maximized,
if one wants to gain an advantage with respect to faint laser pulse systems.
This implies that the collection efficiency of the long wavelength photons
must be particularly optimized through careful alignment of the optical
system, and appropriate selection of the optical components (coating,
numerical aperture). The focal lengths of the lenses located in the three
beams were selected to match the size of their gaussian waist inside the
crystal. We followed the collecting beams in the reverse direction, starting
from the mode field diameter of the fibers and calculating their
transformation through the various components up to the crystal. This mode
matching is essential to obtain a high $\mu $.

To characterize this source, we connect the short wavelength output port to
a Si photon counting detector and the long wavelength one to a gated InGaAs
detector. We obtained a value of approximately 1.1 MHz for the single
counting rate on the Si detector. When monitoring the coincidences in a 2 ns
window using the single channel analyzer of a time-to-amplitude converter,
and taking into account the fact that the quantum detection efficiency of
the InGaAs detector is only 8.5\%, the best value of $\mu $ we obtained was
70\%. Such performance required extremely careful alignment. As far as we
know, it is the best reported. However, a more typical and easily
reproducible value of $\mu $ is $64\%$. It will be used in the rest of the
paper. In order to evaluate the probability to register an accidental
coincidence caused by non-correlated photons, we delayed the coincidence
window by a few nanoseconds. Subtracting the value of the thermal noise of
the InGaAs detector, we measured a value of $\nu $ of $1\%$. We measured the
spectral width of the downconverted photons at 810 nm, and found it to be
smaller than 5nm FWHM.

\subsection{Alice's Interferometer}

In the description of the key distribution principle, it was explained that
Alice and Bob each needed two unbalanced interferometers in order to switch
between two incompatible measurement bases. The path differences of these
interferometers must be matched within a fraction of wavelength, plus or
minus a phase shift of $\pi /2$. They must then be kept stable during the
QKD process. As this condition is very difficult to fulfill, it is
beneficial to devise a system where Alice and Bob have only one
interferometer each. This can for example be achieved by simply inserting in
the interferometers fast phase modulators. However these devices are costly,
and they introduce significant attenuation in the set-up. In addition,
passive state preparation offers superior security, as will be discussed in
Sec. \ref{discus}.

\begin{figure}
\infig{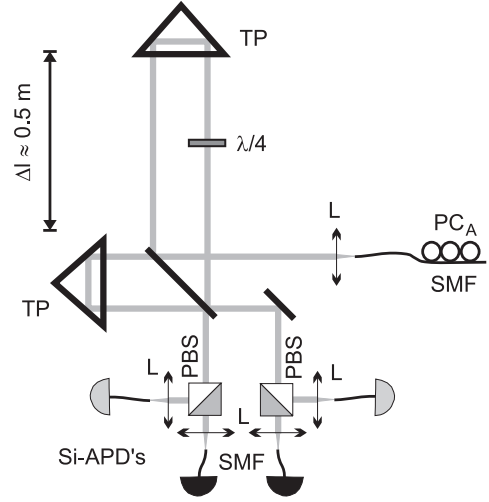}{0.95\columnwidth}
\caption{Schematic diagram of Alice's interfrometer (PC$_{A}$: polarization
controller, SMF: single-mode fibre, L: lens, TP: trombone prism, PBS:
polarizing beamsplitter)}
\label{Fig AliceInterf}
\end{figure}

We devised an elegant alternative. The two interferometers can be
multiplexed in polarization. We add in the long arm of both Mach-Zehnder
interferometers a birefringent element inducing a phase shift of $\pi /2$
between the horizontal and vertical polarization modes. Assuming
constructive interference in one port for vertically polarized light, we
will then observe an equal probability for choosing each output port for
horizontally polarized light. In order to distinguish between both
measurement bases, we also add polarizing beamsplitters separating vertical
and horizontal polarizations between the output ports and the detectors.
When a circularly or 45%
${{}^\circ}$%
-linearly polarized photon enters such a device, it decides upon incidence
on the PBS whether it experienced a phase difference of $\frac{2\pi }{%
\lambda _{A}}\times \Delta L_{A}$ or $\frac{2\pi }{\lambda _{A}}\times
\Delta L_{A}+\frac{\pi }{2}$. Determination of the output port of the PBS
reveals the phase experienced. This principle, offering passive state
preparation, is implemented in Alice's interferometer. Please note that this
polarization multiplexing can also be used with the phase encoding faint
laser pulse scheme introduced by Townsend \cite{townsend93}. When realizing
the interferometers, care has to be taken to keep the interfering events
(Short$_{A}$ - Short$_{B}$, and Long$_{A}$ - Long$_{B}$) as
indistinguishable as possible to maintain high fringe visibility. Because of
the relatively wide spectrum of the down converted photons, chromatic
dispersion may constitute a problem. It should be kept as low as possible in
order to maximize the overlap between both processes. As dispersion in
optical fibers is rather high around 810 nm, we chose to implement Alice's
analyzer with bulk optics, in the form of a folded Mach-Zehnder
interferometer (see Fig. \ref{Fig AliceInterf}). Before launching the
photons into the interferometers, their polarization state is adjusted with
a fiber loops controller. The input port consists of a fiber collimator
(f=11 mm), generating a beam with a diameter of 3 mm. The photons are then
split at a 50/50 hybrid beamsplitting cube (side=25.4mm). We used trombone
prisms (right angle accuracy of $\pm \ 5"$) as reflectors, in order to
simplify alignment. A zero-order quarter wave plate ($\lambda _{0}$ = 800
nm) is inserted in the long arm and vertically aligned to apply the phase
shift on vertical states. A polarizing beamsplitter (side = 11mm, extinction
\mbox{$>$}%
40 dB) is inserted in each output port. Each beam is then focused on the
core of a singlemode fiber (cutoff wavelength
\mbox{$<$}%
780 nm, mode field diameter = 5.5 $\mu $m) using a collimator (NA=0.25, f
=11 mm). The fibers serve as mode filters to yield high fringe visibility.
They are then connected to Si APD photon counting detectors. Although four
such devices are required for complete implementation of the set-up, we had
only two available. When testing the QKD process, we exchanged the fibers to
test all four ports. Both detectors are actively quenched (EG\&G
SPCM-AQR-15FC and SPC-AQ-141-FC). They both have a quantum detection
efficiency of about 50\%, and noise counting rates of the order of 100 Hz.
Whenever a count is registered, the detectors are electronically inhibited
for 500 ns.

The path difference in the interferometers must be larger than the coherence
length of the down-converted photons ($l_{c}\approx 3\cdot 10^{-4}$ m), to
prevent single photon interference. Unfortunately the events are broadened
by the detector's time jitter (of the order of 800 ps FWHM for a coincidence
detection between the first Si APD and an InGaAs APD, and 360 ps FWHM for a
coincidence between the second Si detector and an InGaAs detector, while the
jitter of the InGaAs APD was measured to be 250 ps). The minimum path
difference is thus not limited by the coherence length, but by the width of
the coincidences. In order to keep the overlap between adjacent events below
a few percent, we set the time difference to approximately 3 ns,
corresponding to a round trip path difference of $2\times 0.5=1$ m in air.
This distance should be kept stable within a fraction of a wavelength during
a QKD session. In order to reduce the phase drifts induced by temperature
fluctuations, the interferometer is placed in an insulated box. Moreover,
the temperature is regulated with an accuracy of 0.01$%
{{}^\circ}%
C$. Finally, the mount holding the reflection prism of the long arm is fixed
to the beamsplitter by a glass rod (pure silica), featuring a low linear
expansion coefficient of $5\cdot 10^{-7}$ m$^{-1}\,$K$^{-1}$ (approximately
50 times smaller than that of the aluminum base plate). The length of the
long arm can be varied coarsely by a translation stage with a precision of
approximately $5$ $\mu $m. Fine adjustment is then performed with a
piezoelectric element, featuring a displacement coefficient of about $0.05$ $%
\mu $m$\,/\,V$.

The transmission loss of the interferometer was approximately 9 dB. This
value was very sensitive to the alignment of the reflecting prisms and the
fiber collimators.

\subsection{Bob's Interferometer}

\begin{figure}
\infig{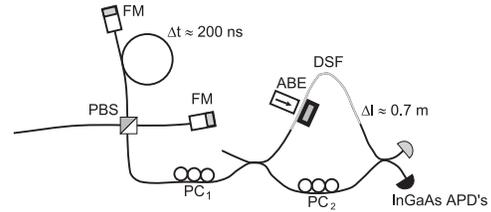}{0.95\columnwidth}
\caption{Schematic diagram of Bob's interferometer (PBS:\ polarizing
beamsplitter, FM: Faraday mirror, PC: polarization controllers, ABE:
adjustable birefringent element, DSF: dispersion shifted fiber).}
\label{FigBobInterf}
\end{figure}

Bob's interferometer is similar to Alice's analyzer, except that it is
implemented with optical fibers (see Fig. \ref{FigBobInterf}). It is
realized with two 3 dB couplers connected to each other. The long arm
consists of DS-fiber with $\lambda _{0}$ close to 1550 nm, in order to avoid
spreading of the photons and maximize the visibility. The path difference is
about 70 cm, corresponding to an optical length of approximately 1 m. A
fiber loop polarization controller is also inserted in this long arm to
ensure identical polarization state transformation for both paths. The
birefringent element used to implement polarization multiplexing consists of
a piezoelectric element applying a variable strain on a 5 mm long uncoated
section of the long arm. This allows tuning the phase difference by
adjusting a continuous voltage. One typically introduces a birefringence of $%
2\pi $ with a voltage of about 50 V, which implies that the adjustment is
not very critical. The exact value depends on the initial strain applied on
the element. In the case of Bob, we separate the two polarizations
corresponding to the measurement bases before injecting the photons in the
interferometer. This information is then transformed into a detection time
information. This is achieved by placing a fiber optic polarizing
beamsplitter (extinction of 20 dB) between the line and the interferometer.
The photons are split according to their polarization and reflected by two
Faraday mirrors, which transform their polarization states into orthogonal
states upon reflection. This ensures that they exit by the port connected to
the interferometer with orthogonal polarizations. While the first arm of
this device measures only 1 meter, the second one is 20 meters longer, so
that a delay of 200 ns is introduced between the two polarization states.
The photon counting detectors are gated twice, and one can infer the
measurement basis, from the detection time bin. As discussed above, after
travelling through the optical fiber line connecting Alice and Bob, the
photons are depolarized. This ensures that each photon will choose randomly
with 50\% probability the basis at the PBS. For example, the degree of
polarization of Bob's photons drops from a value close to 100\% at the
output of the source to only 25\% after an 8.5 km long fiber. However, as
Eve could devise a strategy where she could benefit from forcing detection
of a given qubit in a particular basis, we must introduce a polarizer
aligned at 45%
${{}^\circ}$%
or a polarization scrambler in front of the PBS. As the photons cross the
PBS twice polarized orthogonally, we expect that the imperfections of this
device will only reduce the counting rate, but not introduce errors. The
photons then go through a fiber loops polarization controller ($PC_{1}$) to
align these states with the axes of the variable phase plate. The overall
attenuation of Bob's apparatus is -5.2 dB. It was measured by connecting a
1550 nm LED to the input port of the PBS and by adding the powers measured
at each output ports. This attenuation comes from the insertion loss of the
PBS (1.5 dB), the Faraday mirrors (1 dB) and the couplers (0.5 dB), as well
as the FC/PC connectors. The interferometer is also placed in an insulated
box, where the temperature is kept stable within 0.01%
${{}^\circ}$%
C.

The two detectors connected to the output ports of the interferometer are
EPM 239 AA InGaAs APD's manufactured by Epitaxx. They are mounted on a
measurement stick that is immersed into liquid nitrogen and heated by a
resistor to adjust their temperature to -60%
${{}^\circ}$%
C. The voltage across them is kept below breakdown, except when they are
gated by the application of a 2 ns long and 7.5 V high voltage step \cite
{ribordy98}. The detectors' quantum detection efficiencies are 9.3\% and
9.4\% respectively for a thermal noise probability per gate of $2.8\cdot
10^{-5}$ and $2\cdot 10^{-5}$ (please note that these detectors are
different than the one used to characterize the photon pair source).
Although cooling the detectors to a lower temperature could still further
reduce the thermal noise probability, the lifetime of the trapped charges
yielding afterpulses would increase, so that the overall noise would
actually rise. We checked at -60%
${{}^\circ}$%
C the dependence of the noise probability on the gate repetition frequency.
At 1 MHz, the maximum frequency of our signal generator, a slight increase
was observed. As the minimum time between two subsequent gates is of the
order of 200 ns, and that the repetition frequency does not rise much above
100 kHz, we deduce from this measurement that afterpulses should cause only
limited noise increase in our system.

We discuss the polarization alignment of Bob's interferometer in Sec. \ref
{experim}.

\subsection{Aligning the Interferometers}

The optical path difference of Alice and Bob's interferometers must be
adjusted to be equal within a few wavelengths. This is achieved by
connecting them in series with a scannable Michelson interferometer. Light
from a 1300 nm polarized LED is then injected in this set-up. Because of the
extremely low transmission of the bulk optics interferometer at this
wavelength, the signal is recorded with a passively quenched Germanium
photon counting APD. When scanning the path difference of the Michelson
interferometer, one can register interference fringes when the discrepancy
between the path differences in Alice's and Bob's interferometers is
compensated. This allows measuring $\left| \Delta L_{A}-\Delta L_{B}\right| $
with $\mu m$ accuracy. Because of the chromatic dispersion, this difference
depends on the measurement wavelength. One can compute that at 1550 nm, $%
\Delta L_{B}$ is approximately 400 $\mu $m smaller, in the case of an
interferometer made of DS-fiber, than at 1300 nm. The translation stage in
Alice's interferometer can then be used to adjust $\Delta L_{A}$ and reduce $%
\left| \Delta L_{A}-\Delta L_{B}\right| $ to below a few tens of $\mu $m. At
this point, two-photon interference patterns can be observed when connecting
the photons pairs source to the interferometers. Finally the piezoelectric
element can be used to tune the path difference with an accuracy smaller
than the wavelength.

\subsection{The Classical Channel}

In all QKD systems, a classical channel must be available to perform key
distillation. The experiment reported in this paper features full
implementation of the physical components necessary for QKD. However, we did
not realize the software generating the key from the raw bit sequence. The
classical channel is thus simply used to transport timing information about
the down-converted photons, in order to inform Bob to gate his detectors at
the right time. It consists of a second optical fiber, a 1550 nm DFB laser
at Alice's, and a PIN InGaAs photodiode followed by an amplifier and a
discriminator at Bob's. It features a time jitter of 200 ps, and works with
an attenuation of up to 30 dB. Eve should not be able to gain any
information on the event registered by Alice from the time difference
between the passing photon and classical pulse. The time between the
detection of a single photon and the emission of the classical pulse must
then be equal for the four ports within the time jitter of the photon
counting detectors. This is achieved by adjusting the length of the cables
between the detectors and the electronics. In addition to this timing
signal, we also send on the classical channel information about which
detector registered the count at Alice's. A second pulse, in one of four
time bins, follows thus the synchronization one. Upon detection of a timing
pulse, Bob triggers his detectors and feeds the result he registers along
the decoded information about Alice's detection into a processing unit that
generates several TTL logical signals. Bob can thus keep track of correct
and incorrect events, as well as cases where incompatible bases were used.
For verification purposes, the system also provides false counts in each of
the separate bases. These data are stored on a computer with a digital
counter board (National Instruments PC-TIO-10). In order to implement an
actual key distribution, one must simply remove Alice's information from the
classical channel, by disconnecting one cable. The events are then just
stored by Alice and Bob until key distillation.

\section{Experimental results}

\label{experim}

\subsection{System Adjustment}

Now that the principle of our system and its implementation have been
described, we can present a QKD session. One must first adjust and
characterize the set-up. We assume below that Alice's interferometer is
ready.

The first step is to align the polarization states in Bob's interferometer
with the axes of the birefringent plate. One Faraday mirror is replaced by a
reflectionless termination, so that only one polarization state is sent into
Bob's system. In addition, the short arm of the interferometer, which does
not contain the birefringent element, is opened. A polarized LED at 1550 nm
is injected in the system. One uses then the controller $PC_{1}$ to adjust
the state of polarization, while monitoring it with a polarimeter. The idea
is to find a setting such that applying a voltage on the variable
birefringent element does not modify this state. Once this is done, the
polarization is recorded with the polarimeter and the short arm is
connected. The controller $PC_{2}$ is then used to adjust the transformation
in this arm to bring back the state to the position recorded on the
polarimeter.

\begin{figure}
\infig{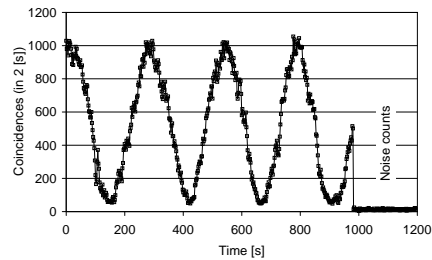}{0.95\columnwidth}
\caption{Typical two photons interference visibility measurement.
Coincidences between a Si-APD at Alice's and an InGaAs-APD at Bob's.}
\label{FigVis}
\end{figure}

The next step is to measure and maximize the visibility of the two-photon
interference fringes. The photon pair source is connected to both
interferometers. One Faraday mirror only is connected at Bob's, so that only
one measurement basis is implemented. It is sufficient to consider one
detector at each side. Alice's detector 1 registers a counting rate of
approximately 100 kHz, with the polarization controller $PC_{A}$ adjusted to
maximize it. In addition, a variable voltage is applied on the piezoelectric
element, varying the length of the long arm in Alice's interferometer. We
used a SRS DS 345 function generator and a piezoelectric controller. The
phase experienced by Alice's photon is thus modulated and two-photons
interference fringes in the coincidences between the detectors can be
recorded (see Fig. \ref{FigVis}). The period is of the order of 4 minutes.
At the end, the delay was modified to measure the noise counts. In the
results presented, we obtained a visibility of $91.8\%\pm 0.8\%$ when
subtracting these noise counts. This value is the same in both bases. Please
note that this measurement basically amounts at performing a Bell inequality
test.

One must then adjust the birefringence in Bob's interferometer, so that the
global phase introduced in both bases equals zero. The second Faraday mirror
is connected, to implement the second basis. The voltage applied on the
birefringent element is slowly tuned until the interference patterns
obtained in each basis are brought in phase. This setting remains stable
during hours.

The last step is to measure the probability for Bob's detectors to produce a
thermal count per gate. We obtain a value of $3.3\cdot 10^{-5}$ and $%
4.4\cdot 10^{-5}$ respectively. The fact that these probabilities are
superior than the figures obtained during the characterization of the
detectors probably comes from the fact that the time between two subsequent
gates is not constant anymore but statistically distributed. Afterpulses may
thus account for this increase. In addition, we have already noticed
significant variations in the performance of InGaAs APD's between
measurements, indicating limited repeatability.

\subsection{Key Distribution}

Now that the system has been tuned and characterized, it is ready for QKD.
Both of Alice's detectors are connected and the polarization controller $%
PC_{A}$ is set so that they each yield the same counting rate. The total
counting rate is approximately of 100 kHz. The voltage applied on the
piezoelectric element varying the length of the long arm of Alice's
interferometer is adjusted manually to minimize the QBER. The key
distribution session can then start and last until the interferometers have
drifted so that the error rate becomes too large. One must then readjust the
voltage on the piezoelectric element. We observed that waiting for two hours
after closing the boxes containing the interferometers ensures higher
stability. We first connected Alice's and Bob's apparatus by a short fiber
of 20 m with essentially no attenuation. Nevertheless, they were located in
two different rooms in order to simulate remote operation.

\begin{figure}
\infig{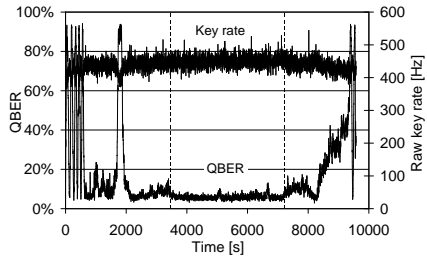}{0.95\columnwidth}
\caption{Key distribution session. The vertical broken lines indicate the
region used to calculate the average QBER. The acquisition time for one data
point was 2 seconds.}
\label{FigSession}
\end{figure}

We obtained a raw key distribution rate (after sifting, but before
distillation) of 450 Hz, and a minimum QBER of $4.7\%\pm 0.3\%$. The whole
key distribution session was defined, somehow arbitrarily, as the period of
time during which the error rate remained below 10\%. It lasted 63 minutes
and allowed the distribution of 1.7 Mbits (see Fig. \ref{FigSession}). The
average error rate, calculated between the vertical dashed lines, was 5.9\%.
It is higher than the minimum because of slight variations in the relative
phase difference in the interferometers induced by temperature drift. Before
and after the key distribution region, fringes were recorded to verify the
interference visibility. It is also possible to estimate the net rate (after
distillation) using the formula presented in \cite{ribordy00}. The fractions
lost during error correction and privacy amplification increase with the
QBER. A value of 178 Hz, readily usable for encryption, can be inferred.

We can apply the formula (\ref{QBERdet}) to (\ref{QBERacc}), and (\ref{Rraw}%
) to verify that these values are consistent with the predictions and to
evaluate the various contributions to the error rate. If we first consider
the equation for the transmission rate, and solve for the detection
efficiency - the quantity exhibiting the most significant uncertainty - we
obtain by setting $\mu =0.64$, $T_{L}=0$ dB, and $T_{B}=-5.2$ dB, an average
quantum detector efficiency $\eta _{D}$ of $8.4\%$. This value is reasonably
close to the expected value of 9.3\%. Considering next the contribution of
the detector noise to the error rate, we can calculate a value of 1\% for $%
QBER_{det}$, by setting $p_{cs}$ to an average value of $3.9\cdot 10^{-5}$
obtained in the last step of the adjustment procedure. From the measured
visibility of 91.8\%, we can infer the contribution $QBER_{opt}$ to be equal
to 4.1\%. Finally, the accidental coincidences contribution to the error
rate can be evaluated to 0.8\% when setting $\nu $ to 1.1\%. These
contributions sum up to a total QBER of 5.9\%, slightly above the minimum
value of the QBER measured ($4.7\%\pm 0.3\%)$. These results are summarized
in Table 1.

We connected then an 8.45 km-long optical fiber spool between Alice and Bob
to verify the behavior of our system. In order to avoid a reduction of the
interference visibility caused by chromatic dispersion spreading, we
selected DS-fiber ($\lambda _{0}=1545$ nm). It featured an overall
attenuation of 4.7 dB. The mode field diameter of this fiber being smaller
than that of the standard fiber used in the source and Bob's interferometer (%
$6\mu m$ instead of $10.5$ $\mu $m), rather high junction losses of 1.3 dB
were obtained at each connection. In addition, the attenuation was 0.25
dB\thinspace /\thinspace km at 1550 nm (measured with an optical time domain
reflectometer). The classical channel was also implemented with an optical
fiber spool whose length was adjusted within 7 cm (360 ps) of that of the
quantum channel.

We first verified that the visibility remained unchanged and obtained a
value of $91.7\%\pm 3.4\%$. This indicates that the use of the DS-fiber
clearly maintains high visibility interference. The measurement of the width
of the coincidence peak between Alice and Bob separated by this DS-fiber
confirms this finding. It is essentially unchanged at 800 ps FWHM, while the
peak broadens to 1.4 ns, yielding substantial overlap of interfering and
non-interfering events (14\% of the non-interfering events within 2ns of the
center of the interference peak), if the standard and DS-fiber are exchanged.

Second, we performed key distribution during 51 minutes at a raw rate of 134
Hz, exchanging 0.41 Mbits. The average QBER was 8.6\% and the minimum QBER $%
6.6\%\pm 0.6\%$. In this case, the net rate is estimated to 32 Hz. On the
one hand, the values of $QBER_{opt}$ (4.1\%) and $QBER_{acc}$ (1.0\%) are
essentially unchanged, as expected. On the other hand, $QBER_{det}$
increased to 3\%. These contributions sum up to 8.1\%, again slightly above
the measured minimum value.

\begin{figure}
\infig{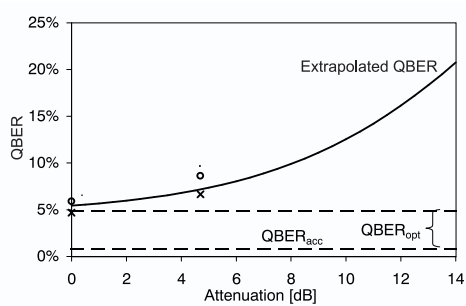}{0.95\columnwidth}
\caption{Experimental values of $QBER_{min}$ (circles) and $QBER_{average}$
(crosses), and extrapolation of the QBER (continuous line). The two
contributions ($QBER_{acc}$ and $QBER_{opt}$) that do not depend on the
distance are also shown (broken lines).}
\label{FigSim}
\end{figure}

One can see on Fig. \ref{FigSim} a graph showing the QBER as a function of
the attenuation of the link between Alice and Bob. It shows the experimental
minimum (circles) and average (diamonds) values obtained with and without
the spool connected. The solid line shows simulated values, with current
InGaAs APD's. The contributions $QBER_{acc}$ and $QBER_{opt}$, independent
on the attenuation , are represented by the dashed lines.

\section{Discussion}

\label{discus}

\subsection{Simulation of the performance with higher attenuation}

We shall now evaluate the potential of this system for application over long
fiber links and compare its performance with two other systems. It is a
straightforward task to extrapolate the results obtained to take into
account the effect of different transmission lines. As discussed above, the $%
QBER_{opt}$ and $QBER_{acc}$ contributions remain unchanged, while $%
QBER_{det}$ increases with the attenuation. Considering Fig. \ref{FigSim},
one can see that, assuming an attenuation coefficient of 0.25 dB\thinspace
/\thinspace km, a QBER of 10\% would be obtained with an attenuation of
approximately 8.5 dB, corresponding to a fiber length of 24 km (0.25
dB\thinspace /\thinspace km and two connections with 1.3 dB). Although these
performances may not seem very good compared for example with the results\
we reported in \cite{ribordy00}, one should remember that the distance is
ultimately limited by the noise performance of the detector. The Epitaxx
detectors used for this experiment show approximately a dark count
probability four times higher than those available at the time of the last
experiment (Fujitsu FPD5W1KS). In addition, the additional losses induced by
the junctions could be reduced by using transition fibers with a slow
variation of their core diameter between the values of standard and
DS-fiber. Alternatively, the system could be completely realized with
DS-fiber. The 8.5 dB attenuation would hence translate into a distance of 34
km.

The accidental coincidences contribution to the error rate could be lowered
in two ways. First, one could reduce the effective width of the gate window
used for the InGaAs APD's. This could be done by feeding the coincidence
signal into a time-to-amplitude converter with a single channel analyzer.
One can estimate that setting the width of this window to one standard
deviation of the coincidence peak (800 ps FWHM) would reduce the accidental
coincidences by a factor of two, while suppressing only one third of the
real coincidences. The ratio of real over accidental coincidences increases
monotically with a reduction of the width of the window. The limit is set by
the reduction of the effective detection efficiency. The dark count
probability would also be reduced by the same factor. The second solution to
reduce this QBER contribution is to decrease the pump power, at the expense
of a reduction of the pair creation rate though. The probability to find an
uncorrelated photon indeed increases with the pump power. This illustrates
why the attenuation in Alice's interferometer does after all matter. If it
is too high, a very high pump power becomes necessary to obtain a given
single counting rate. Nevertheless $QBER_{acc}$ does not really constitute
an important contribution to the error rate, since it is about one percent
and does not grow with distance.

The error contribution of about 4\% due to non-unity visibility is more
serious. This non ideal visibility probably stems from imperfect
polarization alignment in the fiber interferometer, as well as residual
chromatic dispersion. It may also come from a slight difference in the path
difference of Alice's and Bob's interferometers. The two-photon interference
fringes are indeed modulated by a gaussian envelope, whose width is
determined by the coherence length of the down-converted photons. It is
essential to adjust the path differences to be as close as possible to the
maximum of this envelope. However, as the coherence length is rather large,
the top of this envelope is flat and difficult to find. Higher visibilities
(up to 95\%) were indeed obtained but not in a systematically reproducible
way. In practice, we actually observed that it was difficult to tell whether
the visibility improved or not when adjusting the piezoelectric element.
Finally one should also remember that an important feature of $QBER_{opt}$
is that it does not either increase with the distance. However, it would
clearly be valuable to try to improve this visibility.

\subsection{Photon pairs rather than faint laser pulses ?}

It is essential for security reasons when working with faint pulses systems
to keep the fraction of pulses containing more than one photon smaller than
the transmission probability $T_{L}\cdot T_{B}$. If this is not the case,
the spy could use the so-called photon number splitting attack to obtain
substantial information about the key material exchanged (see \cite
{huttner95,yuen96,brassard99} for a discussion of this strategy). She could
indeed measure the number of photons per pulse, and stop all those that do
not contain more than one photon. In turn, when a pulse contains two or more
photons, she splits it and stores one photon, while she dispatches the other
photon to Bob through a lossless medium. Finally, she waits until Alice and
Bob reveal the bases they used to perform her own measurements, and obtains
full information. This potential attack implies that $\mu $ must be reduced,
when the distance is increased. It amplifies the effect of fiber attenuation
on $QBER_{det}$, which limits transmission to even shorter distances.

Our set-up using photon pairs is not vulnerable to this attack. Indeed, even
in the case where two (or more) photon pairs are created within a gate time
of each other, the fact that the state preparation, amounting to the basis
and bit value choices, is made in a passive way ensures that one photon is
not correlated in any way with a photon belonging to another pair. However,
for this to be true, Alice must treat cautiously double detections \cite
{lütkenhausDD}. She cannot simply discard these events, but must assign them
a random value. This increases the error rate, without revealing information
to Eve. When observing two photons in the quantum channel and a pulse in the
classical one, Eve could otherwise deduce that their conjugates took the
same output port at Alice's, yielding a single detection, and are thus
correlated. In practice, because of limited detection efficiency, double
detections are extremely rare. Like the experiment of Tittel \cite{tittel00}%
, our experiment offers thus a superior level of security, which represents
its main advantage over faint laser pulses systems. The two others QKD
experiments performed with photon pairs \cite{naik00,jennewein00} used
active basis switching. Two photons of different pairs are thus invariably
prepared in the same basis. Nevertheless the actual bit value is selected
randomly. In this case, when two photon pairs are emitted simultaneously,
Eve can obtain probabilistic information about the bit value.

To summarize this security issue, we suggest to distinguish three levels.
First, a system could be immune to all attacks, including multiphoton
splitting, like the one presented in this article. In this case, the level
of security is extremely high. Such a system resists attacks with existing,
as well as future technology. Its cost and complexity may however be too
high for real applications. Second, one can consider systems based on faint
pulses. They are immune to existing technology, but would not always resist
multiphoton splitting attacks. However, it is essential here to realize
that, although in principle possible, such an attack would be in practice
incredibly difficult. A natural idea to realize a lossless channel -- one of
the components necessary for such these attacks -- is to use free-space
propagation. However, attenuation in air at 1550 nm is higher than in fibers
(0.64 dB\thinspace /\thinspace km under good visibility \cite{gebbie51}).
Moreover, it depends critically on the atmospheric conditions (in particular
humidity). Diffraction and turbulence induced beam wandering also reduce the
transmission. On the other hand, faint pulses systems offer the advantage to
be reasonably easy to operate and automate. In addition, they could actually
be ready for real applications quickly. Finally, one can look at classical
public key cryptography, which is considered to offer sufficient security,
when implemented with suitable key length. In addition, it is convenient to
apply, as it does not require any dedicated channel, and has been in use for
many years. It however suffers from a major disadvantage. Its security could
indeed be jeopardized overnight by some theoretical advance. In this event,
QKD with faint pulses would constitute the only realistic replacement
technology. In addition, when using public key cryptography, it is essential
to assess the level of computer power that will become available to a
potential eavesdropper during the time the encrypted information bears some
importance. It is indeed also threatened by future developments, while both
types of QKD systems are only vulnerable to technology existing at the time
of the key exchange. QKD with faint pulses may well constitute a compromise
between complexity and security.

A second advantage is that, when Alice detects one photon of a pair, she
knows that a twin photon was also created. This means that we remove the
vacuum component of the faint laser pulses. In principle the probability $%
\mu $ approaches then 1. The correct count probability for a given value of
the attenuation is increased and the contribution $QBER_{det}$ lowered. A
certain QBER will be obtained after a longer distance. It is important to
note that this is beneficial only because detectors are imperfect and
feature noise. If they did not, it would always be possible to compensate
the lower count probability by a larger repetition frequency.

\subsection{Comparison with previous QKD experiments}

We can now compare the performance of the system presented in this paper
with two other set-ups. We first look at the ``plug\&play'' QKD system
presented in \cite{ribordy00}. It features self alignment and highly stable
operation, and was tested by our group over a 22 km long installed optical
fiber in 1998. Our new system in principle allows distribution over a longer
distance. If we take now into account the fact that our source yields a $\mu
$ of only 0.6, we see that the ratio of the detector contributions to the
error rates of both systems is reduced to $QBER_{det}^{P\&P}/QBER_{det}=3/2$%
, instead of $5/2$ when setting $\mu $ to 1. This factor corresponds to an
attenuation of about 1.8 dB, which translates into 7 km of fiber at 1550 nm.
This difference is not really significant. In addition, the ``Plug\&Play''
featured an excellent $QBER_{opt}$ of 0.14\%, and no errors by accidental
coincidences. However, the most important advantage of the system presented
in this paper is clearly the fact that it relies on photon pairs and passive
state preparation, benefitting thus from high security. It does indeed not
offer to Eve at all the possibility to exploit multi-photons pulses for her
attack. We must however admit that the operation of the ``plug\&play'' is
definitely simpler than our system, thanks to its self-alignment feature.
This would also constitute an important parameter when realizing a prototype
to be used by non-physicists. The main difficulty in the manipulation of our
new system comes from the fact that two interferometers must be aligned and
kept stable. The stability problem is of course also encountered with all
the other conventional phase encoding QKD systems \cite{townsend98,hughes00}.

We can also compare it with the system presented by Tittel {\it et al.} in
\cite{tittel00}, who were the first ones to implement QKD with photon pairs
beyond $1$ $\mu $m. They used a pulsed pump laser, whose light passes
through an interferometer, before impinging onto the non-linear crystal and
generating photon pairs. The first measurement basis is implemented exactly
like in the continuous pump system, presented in this paper. No phase change
in the interferometers is required, since the second basis is implemented on
non-interfering events. This implies that the factor $q_{interf}$ has a
value of 1, while the other parameters can in principle have the same value
as that of the continuous pump set-up. This yields a reduction of $%
QBER_{det} $ by a factor 2. On the other hand, the two detectors must be
opened during three time windows, because of the passive basis choice. The
central window corresponds to the first measurement basis using interfering
events, while the two others correspond to the second basis (non-interfering
events). In the system presented here, the detectors are opened only twice.
This implies a $QBER_{det}$ contribution $\frac{3}{2}$ times higher in the
pulsed source system, assuming identical detectors and transmission
attenuation. Overall, this system features a QBERdet contribution 0.75 ($=%
\frac{3}{2}\cdot \frac{1}{2}$) times lower. This factor can be translated
into a gain in distance of about 5 km. Finally, the fact that this pulsed
source system requires alignment and stabilization of three interferometers
(Alice, Bob and the source) however constitutes an additional practical
difficulty.

\section{Conclusion}

In this article, we presented a detailed analysis of quantum key
distribution with entangled states, discussing in particular the noise
sources and practical difficulties associated with these systems. A QKD
system exploiting photon pairs optimized for long distance operation was
tested. We implemented an asymmetrical Franson type experiment for photons
entangled in energy-time and uses a key distribution protocol analogous to
BB84. Passive state preparation, realized by polarization multiplexing of
the interferometers, offers superior security. With Alice and Bob directly
connected, a sifted bit sequence of 1.7 Mbits was distributed at a raw rate
of 450 Hz, and exhibited a QBER of 5.9\%. With an 8.45 km-long fiber between
them, we distributed a sequence of 0.41 Mbits at a raw rate of 134 Hz, and
with an error rate of 8.6\%. We also discussed the level of security offered
by such a system. Finally, we compared the performance obtained with that of
a faint pulse scheme, as well as an alternate one based on entangled photon
pairs.

\begin{acknowledgement}
The Swiss FNRS and OFES as well as the European QuCom project
(IST-1999-10033) have supported this work. The authors would also like to
thank Bruno Huttner for stimulating discussions.
\end{acknowledgement}
\newpage
\begin{table}[tbp] \centering%
%
\begin{tabular}{|c|c|c|c|c|c|c|c|}
Line & Attenuation & Minimum & Average & Raw rate & Duration & Raw key &
Estimated \\
length &  & QBER & QBER &  &  & length & net rate \\
(meters) & (dB) &  &  & (Hz) & (minutes) & (bits) & (Hz) \\ \hline
20 & $\approx 0$ & 4.7\% & 5.9\% & 450 & 63 & 1704118 & 178 \\
8450 & 4.7 & 6.6\% & 8.6\% & 133 & 51 & 407930 & 32
\end{tabular}
\caption{Summary of the performance obtained\label{1}}%
\end{table}%
%

\newpage
\twocolumn

\end{document}